\documentclass[aps,prd,final onecolumn,showpacs]{revtex4}

\usepackage{amssymb}
\usepackage{epsfig,epsf,graphicx}
\usepackage{amsmath}

\newcommand{\be}{\begin{equation}}
\newcommand{\ee}{\end{equation}}

\newcommand{\bee}{\begin{eqnarray}}
\newcommand{\eee}{\end{eqnarray}}

\def\be{\begin{eqnarray} &&}

\def\ee{\end{eqnarray}}
\def\bew{\begin{widetext}}
\def\ew{\end{widetext}}

\usepackage{graphicx}% Include figure files
\usepackage{dcolumn}% Align table columns on decimal point
\usepackage{bm}% bold math

\usepackage{color}
\begin{document}

\title{Cosmological and Particle Physics Constraints on a New Non-Abelian SU(3) Gauge Model for Ordinary/Dark Matter Interaction}

\author{O. Oliveira$^{\dagger,*}$, C. A. Bertulani$^{\ddagger}$, M. S. Hussein$^{\parallel,*}$}
\author{W. de Paula$^*$ and T. Frederico$^*$}

\address{$^\dagger$Departamento de F\'{\i}sica, Universidade de Coimbra, 3004-516 Coimbra, Portugal}
\address{$^\ddagger$ Department of Physics and Astronomy, Texas A\&M University-Commerce,
                                   Commerce TX 75429, USA}
\address{$^\parallel$ Instituto de Estudos Avan\c{c}ados and Instituto de F\'{\i}sica, Universidade de S\~ao Paulo, Caixa Postal 66318,
                                     05314-970 S\~ao Paulo, SP, Brazil}
\address{$^*$Departamento de F\'{\i}sica, Instituto Tecnol\'ogico de Aeron\'autica, DCTA
                               12.228-900, S\~ao Jos\'e dos Campos, SP, Brazil}

\date{\today}

\begin{abstract}
We propose  a mirror model for ordinary and dark matter that assumes a new SU(3) gauge group of transformations, as a natural extension of the Standard Model (SM).
A close study of big bang nucleosynthesis, baryon asymmetries, cosmic microwave background bounds, galaxy 
dynamics,  together with the Standard Model assumptions, help us to set a limit on the mass and width 
of the new gauge boson. The cross section for the elastic scattering of a dark proton by an ordinary proton is estimated
and compare to the WIMP--nucleon experimental upper bounds. It is observed that all experimental
bounds for the various cross sections can be accommodated consistently within the gauge model.
We also suggest a way for direct detection of the new gauge boson via one example of a SM forbidden 
process: $e^+ + p \rightarrow \mu^+ + X$, where $X = \Lambda$ or $\Lambda_c$.
%Further, our predictions can be tested with present high energy experiments at the Large 
%Hadron Collider at CERN and in cosmic rays.
\end{abstract}


%\begin{keyword}
%Dark Matter, Mirror Models, Weakly Interacting Massive Particles
%\end{keyword}

\maketitle

%======================================================
%======================================================
\section{Introduction} 

The mass density ratios computed from the Wilkinson Microwave Anisotropy Probe (WMAP) \cite{WMAP1} data
show that the present day dynamics of the Universe is driven essentially by the Dark Energy (DE), 
see e.g. \cite{Li2011}, Indeed, while $\Omega_{DE} = 0.734\pm0.029$ 
for ordinary baryonic matter, i.e. nuclei and electrons, $\Omega_{b} = 0.0449\pm0.0028$ which is around 
5 times smaller than the corresponding value for dark matter (DM) $\Omega_{DM} = 0.222\pm0.026$.

The nature of dark matter (DM) is a fundamental problem in modern physics. 
Dark matter, see e.g. \cite{Feng2010,bertone05,bertone10,Pato2015},
 is a form of matter that does not interact significantly with ordinary baryonic matter.
Experimental evidence for dark matter comes from the anisotropies of CMB and the dynamics of galaxy clusters.
Elementary particle theory offer scenarios
where new particles such as Weakly Interacting Massive Particles (WIMPs),
Sterile Neutrinos, Axions, Supersymmetric Particles, etc. are possible candidates for DM.
Experimental searches have set limits on the masses and interactions of some of these hypothetical
extra particles \cite{Mavromatos2011,Arina2011,Fornengo2011,DAMA2010,DAMA2010b,CDMS,CoGeNT2011a,CoGeNT2011b,CRESSTII2011,Wagner2011,XENON100,XENON10,ZEPLIN2011,Buen2015,Massey2015}.

A possible scenario for dark matter is the presence of a mirror(s) sector(s) of particles
\cite{Yang,Kobzarev,Pavsic,Foot,Akhmedov} where the mirror sectors are copies of the Standard Model (SM).
If the mirror sectors are no exact copies of the Standard Model, with, e.g. the mirror particles
having different masses and/or couplings than the corresponding SM particles one speaks about 
asymmetric DM models, see e.g. \cite{ADM1,ADM2,ADM3,ADM4}. Anyway, ordinary and mirror particles 
are weakly coupled. Different mirror models provide different echanisms for the coupling between 
ordinary matter and DM.

The current letter describes a mirror model which relies on a symmetry which was so far unexplored. 
Classifying the fundamental matter fields of the Standard Model according to their electric charge leads, 
quite naturally, to an $SU(3)$ symmetry, which can be made local to give dynamics to the interaction.
The model does not requires an \textit{a priori} number of mirror sectors. However, if the dark sectors
are exact copies of the SM, to explain the relative abundance between ordinary and dark matter, five
dark sectors are required. Note that using the quoted values for $\Omega_{DM}$ and
$\Omega_{b}$ it follows that $\Omega_{DM} / \Omega_{b} =4.94 \pm 0.66$; the error on the ratio
was computed assuming gaussian error propagation. Of course, besides the relative abundance the
model should be made compatible with the known cosmological constraints,
with Big Bang Nucleosynthesis (BBN) and with the experimental bounds on the cross sections for
the interaction with ordinary matter.

The gauge model discussed here assumes a new SU(3) symmetry and introduces a 
new weakly interacting massive gauge boson (WIMG) which couples the different sectors and, in this way,
provides the link between dark and ordinary matter. The WIMG, being a massive boson, leaves unchanged the 
long distance properties of the Standard Model and gravity. Further, to describe the interaction between ordinary and dark matter, the model requires only two new parameters, namely the new gauge coupling $g_M$
and the WIMG mass $M$. 

In order to provide mass to the WING, one cannot
rely on the Higgs mechanism which demands the breaking of the underlying gauge symmetry. Instead, 
we introduce a new mechanism for the generation of mass, which requires a scalar in the adjoint representation
with a non-vanishing two point condensate. 
As discussed below, the expectation value of the boson condensate is gauge 
invariant and it turns out that the WIMG mass is proportional to this condensate and, therefore, is also gauge
invariant.

The paper is organized as follows. In section \ref{modelo} the gauge model is described. The problem of the mass
generation and the relation with the SM are explored.

%=====================================================================
%=====================================================================
\section{A Non-Abelian Gauge Model for Dark Matter \label{modelo}} 

At energies much larger than the typical electroweak scale, the Standard Model matter fields behave
like massless particles. At such high energy scales, it is natural to group the matter fields according to their
electric charge
\begin{eqnarray}
 & &  Q_1 = \left( \begin{array}{c} u \\ c \\ t \end{array} \right) , \quad
  Q_2 = \left( \begin{array}{c} d \\ s \\ b \end{array} \right) , \nonumber \\
  & &
  Q_3 = \left( \begin{array}{c} e \\ \mu \\ \tau \end{array} \right) , \quad
  Q_4 = \left( \begin{array}{c} \nu_e \\ \nu_\mu \\ \nu_\tau \end{array} \right) \, .
  \label{matter}
\end{eqnarray}
From the point of view of the electric charge, each multiplet can be rotated to produce a physically equivalent multiplet. 
This property suggests a new $SU(3)$ group of transformations, where the $Q_f$'s belong to the 
fundamental representation, which is broken at the electroweak scale. At the level of the lagrangean density
the breaking of the new symmetry shows up in the mass matrix whose origin is linked with the Higgs mechanism
within the SM. Although, in the SM the fermion fields are chiral fields, in the definition of the $Q$ chirality we will
explore the possibility of having non-chiral fields.
Therefore, in the model described below, when exploring the implications for 
particle physics phenomenology, we will consider two cases in what concerns the chiral properties of $Q_i$: 
(i) the matter fields in $Q_f$ do not include a chiral projector in their definition, 
named non-chiral theory below; (ii) the fields in $Q_f$ are all left-handed, called chiral theory below, 
and a $\gamma_L = ( 1 - \gamma_5 )/2$ should be attached to each field in (\ref{matter}). 

The set of fields $Q_f$ in (\ref{matter}) are the ordinary SM matter. In the following, we will assume that
DM has a similar structure as observed for ordinary matter. Each DM sector has 4 multiplets which mimic 
(\ref{matter}) and each sector has its own copy of the SM. For the moment, the number of DM sectors
needs not to be fixed. Given that DM does not seem to couple with the photon, it will also be assumed that
each sector is blind to the remaining sectors. Therefore, each sector has its own copy of the SM, with the
the corresponding electroweak sectors bosons coupling only within the sector that they are associated with.
Note that we are not assuming that each sector is an exact copy of the known ordinary family sector. 
Indeed, looking at the SM and its family structure what is observed is that the masses differ,
not only within each family, but also when one considers different families.
In principle, a similar situation can occur here, with the various DM families having different couplings and 
different masses than those found for the SM. 

A dynamics between the various families can be defined if the new global $SU(3)$ symmetry is made local. This
requires the introduction of a new gauge boson, the WIMG, coupling all sectors. Note that it will be assumed that
the different sectors will be coupled via WIMG exchange and by the gravitational interaction. 
In our cold Universe, for a sufficient high WIMG mass or a sufficient small coupling constant, 
the connection between the different sectors is suppressed or essentially vanishing or is
provided by gravitational coupling.

The gauge model includes the WIMG field $M^a_\mu$, the matter fields $Q_f$, where $f$ is a flavor index,
and a real scalar field $\phi^a$ belonging to the adjoint representation of the $SU(3)$ group. 
The scalar field is required to provide a mass to $M^a_\mu$ and, in this way, it ensures
that the WIMG interaction is short distance.
% and the
%In this way, 
%new contributions to the long distance nature of the gravitational force is avoided.

The Lagrangian  for the gauge theory reads
\begin{eqnarray}
 \mathcal{L}  & = &  - \frac{1}{4} F^a_{\mu\nu} F^{a \, \mu\nu} ~ + ~
 \sum_f \overline Q_f \, i \gamma^\mu D_\mu \, Q_f
 ~ + ~ \nonumber \\ 
 & & + ~ \frac{1}{2} \left( D^\mu \phi^a \right) \left( D_\mu \phi^a \right) - V_{oct}( \phi^a \phi^a )
% ~ + ~ \mathcal{L}_{GF} ~ + ~ \mathcal{L}_{gh}
 \label{lagrangeano}
\end{eqnarray}
where $D_\mu = \partial_\mu + i g_M T^a M^a_\mu$  is the covariant derivative,
$T^a$ stands for the generators of $SU(3)$ group, $m_f$ the current quark mass matrix and
$V_{oct}$ is the potential energy associated with $\phi^a$. 
The second term in (\ref{lagrangeano}) includes a sum over all families of matter, ordinary and dark matter, 
and, within each family, over the four multiplets. We have omitted in $\mathcal{L}$ the terms defining the
SM for each family and the terms associated with the quantization of the theory. 

%========================================================================
%========================================================================
\subsection{Scalar Fields and WIMG Mass}

In the model, we require the WIMG to be a massive gauge boson to leave the long distance properties of the
gravitational interaction unchanged. In order to generate a mass to $M^a_\mu$ keeping gauge
invariance, one has to rely on scalar fields. 

The Higgs mechanism leaves a number of components of $M^a_\mu$ massless and, 
to keep the long distance forces unchanged, the Higgs mechanism must be excluded as a way to 
give mass to the gauge fields. 

In \cite{Oliveira2011}, the authors propose a mechanism for mass generation via the introduction 
of a scalar condensate which complies with gauge invariance and provides the same mass for all
the components of the gauge field. In the following, we will assume that the WIMG acquires mass through 
this mechanism. For completeness we will provide the details of the mass generation mechanism.

The kinetic term associated with the scalar field accommodates a mass term for the WIMG field. The gauge field
mass term is associated with the operator
\begin{equation}
  \frac{1}{2} \, g^2_M \, \phi^c (T^a T^b)_{cd} \phi^d M^a_\mu M^{b \, \mu} \, .
\end{equation}
The scalar field cannot acquire a vacuum expectation value without breaking gauge invariance. However,
to generate a mass for the WIMG it is sufficient to assume a non-vanishing boson condensate
$\langle \phi ^a \phi^b \rangle$. The origin of this condensate can be associated with local fluctuations of
the scalar field. 

If the dynamics of the scalar field is such that
\begin{equation}
\langle \phi ^a \rangle = 0 \qquad\mbox{ and }\qquad \langle \phi^a \phi^b \rangle = v^2 \delta^{ab} \, ,
\label{condensado}
\end{equation}
given that for the adjoint representation $\mbox{tr} (  T^a T^b )= 3 \, \delta^{ab}$, it follws that
the square of the WIMG mass reads
\begin{equation}
 M^2  = 3 \, g^2_M   v^2 \, .
 \label{WIMG_mass}
\end{equation} 
Note that $v^2$ and, therefore,  the WIMG mass are gauge invariant. The proof of gauge invariance follows 
directly from the transformations properties of $\phi^a$. 

The vacuum expectation values required to give mass to the WIMG  (\ref{condensado}) minimize the type of
potential used in the Higgs mechanism, 
\begin{equation}
   V(\phi) = \lambda \left( \phi^a \phi^a -  8 v^2 \right)^2 \, ,
\end{equation}   
but without breaking the $SU(3)$ symmetry, as would happen in the case of the standard Higgs mechanism.
The relations derived above are valid at the tree level and one should investigate how they are changed
by the quantum corrections. 
However, if the theory has a perturbation solution, as is assumed here, at least in lowest
order the mass and vacuum expectation values relations should hold.

The WIMG mass (\ref{WIMG_mass}) is proportional to the effective gauge coupling $g_M$. One expects
to have a small enough $g_M$. This does not imply necessarily that $M$ is small. The precise value of $M$
depends on the relative values of $v$ and $g_M$.

%========================================================================
%========================================================================
\subsection{Other Symmetries of $\mathcal{L}$}

The Lagrangian density $\mathcal{L}$ has several symmetries, both discrete and continuum,
besides the local SU(3) gauge invariance.
In the present work we are mainly interested in exploring the implications of (\ref{lagrangeano}) 
for DM and will not explore the full structure "hidden" in $\mathcal{L}$. 
However, lets us comment one some of these symmetries.

The sectors in $\mathcal{L}$ are copies of the ordinary matter. Therefore, it is possible to assign
a baryonic and a leptonic number for each sector and, within each family, they are conserved.

The lagrangean density (\ref{lagrangeano}) is invariant under rotations of $Q_f$. This $SU(4)$ flavor
like symmetry per sector mixes leptons and quarks and is broken explicitly by the mass terms. 
In the case of symmetric DM, where all sectors are  copies of the ordinary matter sector, this
invariance defines a larger group, i.e. one can rotate the $Q$'s not only within each family but also
between sectors.

In the SM the generation of the mass for the fermion fields occurs, within in the electroweak sector. 
The diagonalization of the mass matrix requires a rotation of the fields by a unitary transformation
over the up and down type of flavors. 
In (\ref{lagrangeano}), such a rotation introduces an unitary flavor matrix
associated with the WIMG vertex. For example, in this way, the model can accommodate neutrino mixing.

The implications of the $\mathcal{L}$ symmetries will be the addressed in future publications \cite{futuro}.

%=========================================================
%=========================================================
\section{WIMG Properties and Cosmological Constraints}

In this section we described how the cosmological constraints can be accommodated within
the non-abelian gauge model.

%=========================================================
%=========================================================
\subsection{Big Bang Nucleosynthesis and Baryon Asymmetries} 

The gauge model summarized in (\ref{lagrangeano}) has new relativistic degrees of freedom that can
increase the expansion rate of the early Universe \cite{Berezhiani1996} and affect  
big bang nucleosynthesis (BBN) \cite{Hoyle}.

After inflation, the temperature for the thermal baths associated with each particle species
is not necessarily the same \cite{Berezhiani1996}. It depends on the various possible reactions enabling equilibria
and on the Universe thermal history. Let us start discuss the simplest possible picture where all the
dark sectors have the same temperature, different from the ordinary matter thermal bath, i.e. we are assuming 
that asymmetric reheating takes place after inflation as in \cite{Kolb,Berezhiani2001,Ciarcelluti}.

The number of possible new particles contributing to the radiation density during the BBN epoch are
constrained by the $^4$He primordial abundance and the baryon-to-photon ratio
$\eta = n_b/n_\gamma$, where $n_b$ is the baryon density and $n_\gamma$ the photon density in the Universe
\cite{BBNLimit}.
For a radiation dominated Universe at very high temperatures, neglecting the particles masses,
the energy and entropy densities are given by \cite{pdg2010}
\begin{equation}
 \rho(T)=\frac{\pi^{2}}{30} \, g_*(T) \,T^4 \quad\mbox{and}\quad
 s(T)=\frac{2\pi^{2}}{45} \, g_s(T)\, T^3 ,
 \label{energy_entropy}
\end{equation}
where
\begin{equation}
g_*(T)=\sum_B g_B \left(\frac{T_{B}}{T}\right)^4 + \frac{7}{8} \sum_F g_F \left(\frac{T_{F}}{T}\right)^4
\end{equation}
and
\begin{equation}
g_s(T)=\sum_B g_B \left(\frac{T_{B}}{T}\right)^3 + \frac{7}{8} \sum_F g_F \left(\frac{T_{F}}{T}\right)^3,
\end{equation}
are the effective number of degrees of freedom during nucleosynthesis, $g_{B(F)}$ is the number of degrees of
freedom of the boson (fermion) species $B(F)$, $T_{B(F)}$ is the temperature of the thermal bath of species 
$B(F)$ and $T$ the temperature of the photon thermal bath.

In our case, we consider that the ordinary and dark sectors are decoupled, just after reheating, with different
temperatures: $T$ for ordinary matter and $T'$ for the dark sectors. For the dark sectors,
the energy $\rho'(T')$ and entropy $s'(T')$ densities are given as in (\ref{energy_entropy})
after replacing $g_*(T) \rightarrow g'_*(T')$ and $g_s(T) \rightarrow g'_s(T')$, i.e.
the effective number of degrees of freedom in the dark sector, and replacing $T$ by $T'$.
The entropy in each sector is separately conserved during the Universe evolution,
which leads that $x=(s'/s)^{1/3}$ is time independent. Assuming the same relativistic particle
content for each sector of the modern universe, one has  $g_s(T_0) = g_s^\prime(T'_0)$ and
it follows that  $x=T'/T$.

For a radiation dominated era, the Friedman equation is
\begin{equation}
H(t)=\sqrt{\left(8\pi/3 c^2\right) \, G_{N} \, \bar{\rho}},
\end{equation}
where the total energy density is given by $\bar{\rho} = \rho\, + \, N_{DM} \, \rho'$, where
$N_{DM}$ is the number of dark sectors. From the expression for $\rho'$,
it follows
\begin{equation}
H(t)=1.66 \, \sqrt{\bar{g}_{*}(T)} \, \frac{T^2}{M_{Pl}},
\end{equation}
where
\begin{equation}
\bar{g}_{*}(T) = g_{*} (T) \left( 1+ N_{DM} \, a \, x^4 \right)
\end{equation}
and $M_{Pl}$ is the Planck mass.
The parameter $a = \left(g'_{*}/g_{*}\right) \left(g_{s}/g'_{s}\right)^{4/3}\sim 1$, unless $T'/T$ is very
small \cite{Berezhiani1996}. At the nucleosynthesis temperature scale of about 1 MeV, the
relativistic degrees of freedom (photons, electrons, positrons and neutrinos)
are in a quasi-equilibrium state
and $g_{*}(T)|_{T=1MeV} = 10.75$.
The extra dark particles
change $g_{*}$ to $\bar{g}_{*} = g_{*} \left(1 + N_{DM} \, x^4 \right)$. The bounds due to the relative
abundances of the light element isotopes ($^{4}$He, $^{3}$He, D and $^{7}$Li) are usually
written in terms of the equivalent number of massless neutrinos during nucleosynthesis:
$3.46 <  N_{\nu} < 5.2$ \cite{WMAP-cosm}.
The extra degrees of freedom introduced by the dark sectors lead to
$\Delta g_{*}=\bar{g}_{*}-g_{*} = 1.75 \, \Delta N_\nu < 3.85$, where $\Delta N_\nu$ is the variation
in equivalent number of neutrinos, and  $T' / T < 0.78 / N^{1/4}_{DM}$ to reconcile the gauge model
with the BBN data.
If $N_{DM} = 5$, as required by to explain the observed ratio $\Omega_{DM} / \Omega_b$, then
$T' /T < 0.52$. In conclusion, the asymmetric reheating mechanism leads always to dark universes
which are colder than our one universe.

The baryon asymmetry is parameterized by the baryon-to-photon ratio $\eta$.
The density number of photons $n_{\gamma}$ is proportional to T$^{3}$ and, therefore,
one can write the density number of dark-photons as $n'_{\gamma}=x^3 n_{\gamma}$. The
ratio of dark-baryons to ordinary-baryons is given by $\beta=\Omega'_{B}/\Omega_{B} = x^3 \eta'/\eta$
\cite{Berezhiani2001}. The bounds from the BBN on $x = T' / T$ imply that the baryon asymmetry in the
dark sector is greater than in the ordinary one. Indeed, using the upper bound $x\sim 0.78/N^{1/4}_{DM}$ 
and assuming that each sector contributes equally to the Universe's energy
density $\beta\sim 1$, we obtain $\eta' \sim 2.1 \, N^{3/4}_{DM}\eta$. For the special where
$N_{DM} = 5$ it follows that $\eta'\sim 7\eta$. Asymmetric Dark Matter models, see e.g.
\cite{ADM1,ADM2,ADM3,ADM4}, give similar results for the baryon asymmetry.
%
%OLD VERSION*******************************************************************************
%Indeed, using the upper bound $x\sim 0.5$ and for the most
%symmetric assumption ($N_d = 5$), where each sector contributes equally to the Universe's energy
%density $\beta\sim 1$, we obtain $\eta'\sim 8\eta$. Asymmetric Dark Matter models, see e.g.
%\cite{ADM1,ADM2,ADM3,ADM4}, give similar results for the baryon asymmetry.

In principle, the presence of mirror baryon dark matter (MBDM) could give some effect on the CMB power 
spectrum. The reason is that the acoustic oscillations of MBDM could be transmitted to the ordinary baryons. 
In Ref. \cite{Berezhiani2001} this effect was analyzed and their conclusion is that to obtain
an observable effect in CMB data it is necessary to have a ratio of temperatures $T'/T\geq 0.6$. 
This bound combined with the BBN analysis provides a lower bound for the number of dark sectors:
$0.35 < N_{DM}$.

%=========================================================
%=========================================================
\subsection{Galaxy Dynamics and Long Range Interactions} 

Galaxy dynamics provide further constraints on DM, see e.g.
\cite{Mohapatra2002,Ackerman2009}.
In the gauge model there is no direct coupling between the photon and its dark brothers.
Further, it is assumed that the different sectors behave as the ordinary matter family.
It seems natural that the galaxy halos are neutral relative to the $U(1)$'s within each
sector.
The observed dark matter halos suggest that DM are effectively
collisionless and demand an upper bound in the cross section of DM-DM interactions
\cite{Wandelt2000,Escude2000,Dave2001,Yoshida2000}.
The $T'/T$ bound estimated from BBN complies with such a statement.
A typical cross section is given by
$\sigma \approx ( g^2 T / \Lambda^2)^2$, where $g$ is the interaction coupling constant,
$T$ is the temperature and $\Lambda$ a typical mass scale of the interaction.
If the dark sectors are copies of the ordinary matter sector, i.e. $g$ and $\Lambda$ are of the same
order of magnitude, one can write $\sigma'/\sigma = (T'/T)^2$, where
$\sigma'$ ($\sigma$) is the cross section for the dark (ordinary) family. The temperature
bound from BBN 
implies that $\sigma' / \sigma < 0.61/\sqrt{N_{DM}}$
and, as long as $T' / T$ is sufficiently small, DM becomes effectively
collisionless.

%=============================================================
%=============================================================
\section{Particle Physics Phenomenology} 

For the first hadronic ordinary family, the WIMG$\!\!$--quark interaction part of the Lagrangian is
\begin{widetext}
\begin{eqnarray}
 \mathcal{L}_{Mq} & = &
  \frac{g_M}{2}?\left[ \overline u \, \gamma^\mu \Gamma \, c \right] \left( M^1_\mu - i M^2_\mu \right) +
  \frac{g_M}{2}?\left[ \overline u \, \gamma^\mu \Gamma \, t \right]  \left( M^4_\mu - i M^5_\mu \right) +
   \frac{g_M}{2}?\left[ \overline c \, \gamma^\mu \Gamma \, t \right] \left( M^6_\mu - i M^7_\mu \right) 
  \nonumber \\
  & & 
  +
   \frac{g_M}{2}?\left[ \overline u \, \gamma^\mu \Gamma \, u \right] \left( M^3_\mu + \frac{1}{\sqrt{3}} M^8_\mu \right) +
  \frac{g_M}{2}?\left[ \overline c \, \gamma^\mu \Gamma \, c \right] \left( - M^3_\mu + \frac{1}{\sqrt{3}} M^8_\mu \right) +
  \frac{g_M}{2}?\left[ \overline t \, \gamma^\mu \Gamma \, t \right] \left( - \frac{2}{\sqrt{3}} M^8_\mu \right)  
  \nonumber \\
  & & + \,  h.c. \, ,
  \label{L_q_mulato}
\end{eqnarray}
\end{widetext}
where
$\Gamma = I$ for the non-chiral theory and $\Gamma = \gamma_L$ for the chiral theory.
The remaining families have similar patterns of interactions. The new vertices give rise to flavor changing
type of processes within the same family. Given that the WIMG
has no electrical charge, it seems that it can give rise to flavor changing neutral processes which are, at most,
suppressed by $\sim g^2_M/M^2$. However, the flavor structure of (\ref{L_q_mulato}) and given that
the WIMG propagator is flavor diagonal, the $S-$matrix element for these processes vanishes.
For example, as discussed below, the WIMG vertices give no contributions to the lepton flavor violation
processes reported in the particle data book \cite{pdg2010}. In this sense, the WIMG interaction is
compatible with the GIM (Glashow-Iliopoulos-Maiani) mechanism and the flavor changing
neutral currents should remain suppressed at high energies. However, the interaction Lagrangian (\ref{L_q_mulato})
allows lepton family number violation, see e.g. figures \ref{mudecay} and \ref{flavorx}.

%===============================================================================
%===============================================================================
\subsection{Lepton Anomalous Magnetic Moment} 

\begin{figure}[b]
\centering
\epsfig{file=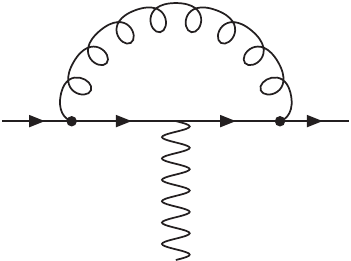,angle=0, width=4cm}
\caption{Lepton-photon vertex correction by WIMG exchange.} \label{photonvertex}
\end{figure}

The new gauge boson  provide corrections to the lepton-photon vertex which
contribute to the lepton anomalous magnetic moment as depicted in fig.\ref{photonvertex}.
The new contributions to $(g-2)/2$ due to the WIMG are UV-finite and read
\begin{equation}
 a_{e,\mu} =  \frac{g^2_M}{16 \pi^2} \left( \frac{m_{e,\mu}}{M} \right)^2\left(\frac53-
\frac34\frac{m_\tau}{m_{e,\mu}}\right) 
\label{eq:g2_nonchiral0}
\end{equation}
and
\begin{equation}
 a_\tau =  \frac{7\, g^2_M}{96 \pi^2} \left( \frac{m_\tau}{M} \right)^2 \ ,
\label{eq:g2_nonchiral1}
\end{equation}
for the non-chiral theory and
\begin{equation}
 a_l = \frac{5\, g^2_M}{96 \pi^2} \left( \frac{m_l}{M} \right)^2  \ ,
\label{eq:g2_chiral}
\end{equation}
where $l=(e,\mu,\tau)$, if the particle in the multiplets are left-handed.
In (\ref{eq:g2_nonchiral0}), (\ref{eq:g2_nonchiral1}) and (\ref{eq:g2_chiral}) only
the leading contributions in $m^2_l / M^2$, where $m_l$ is the lepton mass and $M$ the WIMG mass, are
taken into account.

The particle data group \cite{pdg2010} quotes the following values for the anomalous magnetic moment
$a_l = ( g - 2 )_l / 2 = \left( 1159.65218073 \pm 0.00000028 \right) \times 10^{-6}$ for $l = e$,
$a_\mu =  \left( 11659208.9 \pm 5.4 \pm 3.3 \right) \times 10^{-10}$ 
and $a_\tau  >  -0.052$ and $ < 0.013$. 
For the muon there is a $3.2\,  \sigma$ difference between the experimental value $a^{exp}_\mu$
and the standard model prediction $a^{SM}_\mu$ which is of
$  \Delta a_\mu = a^{exp}_\mu  - a^{SM}_\mu = 255 (63) (49) \times 10^{-11}$.

For the non-chiral theory the WIMG contribution to the lepton anomalous magnetic moment is, for the electron
and for the $\mu$, negative due to the $\tau$ loop correction to the vertex. Therefore, in the non-chiral theory, the
WIMG cannot explain $\Delta a_\mu$ and $a_{e, \mu}$ should be, at most, of the order of the experimental error.
This provides the constraints
\begin{equation}
    \frac{g^2_M}{M^2} \leq 6.50 \times 10^{-14} \mbox{ MeV}^{-2}
\end{equation}
or
\begin{equation}
    \frac{g^2_M}{M^2} \leq 8.14 \times 10^{-13} \mbox{ MeV}^{-2}
\end{equation}
if one uses $a_e$ or $a_\mu$; for $a_\mu$ the errors reported in \cite{pdg2010} were added in quadrature.
In the above calculation we used $m_e = 0.511$ MeV and $m_\mu = 105.658$ MeV.
These bounds can be rewritten in terms of the WIMG mass as
$ M \geq g_M \times 3.9$ TeV  and $M \geq g_M \times 1.1$ TeV,
respectively. 

The WIMG contribution to the $\tau$ anomalous magnetic momenta should be $ \sim 1.5 \times 10^{-9}$ or smaller.
For the chiral theory, the  WIMG contribution to $(g-2)/2$ should comply with the above results and can be, at most, of the order
of the muon anomaly $\Delta a_\mu$, i.e.
$ a_\mu \leq 255 (63) (49) \times 10^{-11} $, therefore
\begin{equation}
   \frac{g^2_M}{M^2} \leq 4.33 \times 10^{-11} \mbox{ MeV}^{-2}
\end{equation}
and the WIMG mass should obey $ M  \geq g_M \times 0.152$  TeV.
For the non-chiral theory, the choice of $\Delta a_\mu$
to define the WIMG mass complies with the experimental error for the electron
and tau. Indeed, from (\ref{eq:g2_chiral}) it follows that the contribution of
the new gauge bosons to the electron/tau magnetic moment is
$a_l = (m^2_l / m^2_\mu) \, a_\mu$.
These scaling laws give an $a_e = 6.0 \times 10^{-14}$  and $a_\tau = 7.2 \times 10^{-7}$
which are smaller than the experimental error.

%==================================================================
%==================================================================
\subsection{Lepton Flavor Violation Decays} 

The Lagrangian (\ref{L_q_mulato}) allows for flavor changing processes within the same family.
The WIMG propagator is flavor diagonal, therefore only those processes where the propagator links
the same type of vertices at both ends can have a non-vanishing $S-$matrix. The following lepton family
number violating decays
\begin{eqnarray}
&&\mu^- \longrightarrow
e^- \nu_e \overline\nu_\mu , \ \
\mu^- \longrightarrow e^- e^+ e^- , \ \
\tau^- \longrightarrow e^- e^+ e^-, \nonumber
\\
&&\tau^- \longrightarrow  e^- \mu^+ \mu^-, \ \
\tau^-\longrightarrow \mu^- e^+ e^-, \ \
\tau^- \longrightarrow \mu^+ e^+ e^-,  \nonumber
\end{eqnarray}
are forbidden within the model because they require different vertices connected
by the WIMG propagator. On the other hand, the quark-WIMG vertex structure gives a
vanishing $S-$matrix for $\mu^- \rightarrow e^- \nu_e \overline \nu_\mu$. Processes with photons,
such as, $\mu^- \rightarrow e^- \gamma$ or $\mu^- \rightarrow e^- \gamma\gamma$ can only occur via loops
and are highly suppressed at low energies. The same arguments apply to
$B_d \rightarrow e^-\tau^+$ and $B^0_s \rightarrow \mu^+\tau^-$ which are forbidden in the model.
It turns out that the gauge theory described here complies with the lepton flavor violation bounds
reported in the particle data book.

%
%\begin{figure}[t]
%\centering
%\epsfig{file=fig2.eps,angle=0, width=4cm}
%\hspace{1cm}
%\epsfig{file=fig3.eps,angle=0, width=5cm}
%\vspace{.5cm}
%\caption{Muon decay to $(e^-,\nu_\mu,\overline \nu_e)$ by WIMG exchange process (\textit{Left}).
%              $D^0$ decay to $(\mu^+,e^-)$ by WIMG exchange process (\text{Right}).} \label{mudecay}
%\end{figure}
%

%========================================================
%========================================================
\subsection{Muon Beta Decay}

\begin{figure}[t]
\centering
\epsfig{file=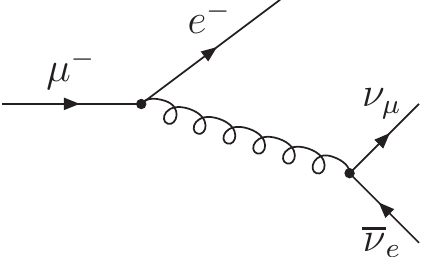,angle=0, width=4cm}
\caption{Muon decay %************** to $(e^-,\nu_\mu,\overline \nu_e)$ 
              by WIMG exchange process.} \label{mudecay}
\end{figure}

The WIMG can also contribute to the leptonic decays of the $\mu$, the $D$'s and the $B$'s mesons.
We start by computing the main muonic decay channel $\mu^- \rightarrow e^- \nu_\mu \overline\nu_e$ as
shown in fig. \ref{mudecay}. The $S-$matrix
gets a contribution from $W$ exchange and WIMG exchange. To leading order in $M_W$ and $M$, for the
chiral theory the matrix element for the transition reads
\begin{equation}
  \overline{ \left(  i \mathcal{M} \right)^2 } = 64 \, G^4_F \, \left[ 1 - \frac{1}{2 \sqrt{2}}
\frac{ g^2_M /M^2}{G_F} \right] \,
  \left(p_\mu \cdot p_{\overline\nu_e} \right) \,
  \left(p_e \cdot p_{\nu_\mu} \right) \, .
\end{equation}
For the chiral theory an extra factor of $1/2$ should multiply the $g^2_M/M^2$.
The new contribution modifies the Fermi coupling constant as follows
\begin{eqnarray}
   G_F & \longrightarrow & G_F \, \left[ 1 - \frac{1}{2 \sqrt{2}} \frac{ g^2_M /M^2}{G_F} \right]^{1/4}  
   \nonumber \\
   & &
   \approx G_F \, \left[ 1 - \frac{1}{8 \sqrt{2}} \frac{ g^2_M /M^2}{G_F} \right]  \, ,
\end{eqnarray}
with $G_F = 1.16637(1) \times 10^{-5}$ GeV$^{-2}$  \cite{pdg2010}.
Requiring that the WIMG contribution to be of order of
the error on $G_F$ or smaller, then
\begin{displaymath}
  \frac{1}{8 \sqrt{2}} \frac{ g^2_M}{ M^2}  \le 1.0 \times 10^{-10} \, \mbox{ GeV}^{-2} 
\end{displaymath}
or
\begin{equation}
  \frac{ g^2_M}{ M^2}  \le 1.13 \times 10^{-9} \, \mbox{ GeV}^{-2} \, .
  \label{bound_tramado}
\end{equation}
The relative error on the muon decay width produces a slightly larger $g^2_M/M^2$ lower limit.
The corresponding mass bound is 
$M \geq g_M \times 28$ TeV.
Assuming that $g_M \approx e = \sqrt{4 \pi \alpha} = 0.30$, we have a lower bound the WIMG mass
of $\approx 9$ TeV which complies with both the $\beta$-decay, i.e. the error on the Fermi coupling constant,
and the muon decay. These mass bounds are more restrictive than the bounds from the
anomalous magnetic moment.

%===================================================================
%===================================================================
\subsection{$B$ and $D$ Leptonic Decays}?

\begin{figure}[t]
\centering
\epsfig{file=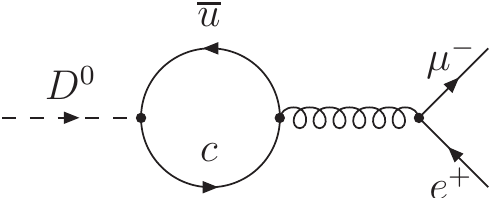,angle=0, width=5cm}
\caption{$D^0$ decay %********************
              to $(\mu^+,e^-)$ by WIMG exchange process.} \label{D0decay}
\end{figure}
The  WIMG vertices can give rise to the following leptonic decays
\begin{displaymath}
 D^0 (c \overline u) \longrightarrow \mu^- e^+, ~ 
 B^0 (b \overline d) \longrightarrow \tau^- e^+, ~
 B^0_s (s \overline b) \longrightarrow \tau^- \mu^+
\end{displaymath}
and complex conjugate decays. The widths can be computed using the relation
$\langle 0 | ~ \overline q \, \gamma^\mu \gamma_5 \, q^\prime ~ | (q \overline q^\prime ) \rangle = i \, f \, P^\mu$,
where $| (q \overline q^\prime ) \rangle$ stands for the meson state composed of
quarks $q \overline q^\prime$, $f$ is the meson decay constant and $P$ the four-momentum of the meson.
Note that the above decays are possible only within the chiral theory. The heavy meson leptonic decay width
is calculated by evaluating the amplitude shown in fig. \ref{D0decay}, exemplified for $D^0\to\mu^-e^+$, which
in the general case is given by:
\begin{equation}
  \Gamma = \frac{1}{256 \, \pi} \, \frac{g^4_M}{M^4} \, f^2 \, m^2_l \, m_m \, \left( 1 - \frac{m^2_l}{m^2_m} \right)^2 \, ,
  \label{largura_mesao}
\end{equation}
where we have assumed that the lightest lepton is massless, $m_l$ is the mass of the heavier lepton and
$m_m$ the mass of the meson state.

For the decay $D^0 \rightarrow \mu^- e^+$, using a $m_{D^0} = 1.864$ GeV and $f_{D^0} = 0.206$ GeV,
the bound
(\ref{bound_tramado}) implies for the corresponding branching ratio
\begin{equation}
  Br(D^0 \rightarrow \mu^- e^+) < 8.7 \times 10^{-13} \,
   \label{bound_D0}
\end{equation}
to be compared with the experimental limit \cite{pdg2010} of
$  Br(D^0 \rightarrow \mu^- e^+) < 2.6 \times 10^{-7}$.
For the decay $B^0 \rightarrow \tau^- e^+$, using a $m_{B^0} = 5.279$ GeV \cite{pdg2010}
and $f_{B^0} = 0.22$ GeV \cite{Mohanta2010}, the bound (\ref{bound_tramado}) gives a
\begin{equation}
  Br(B^0 \rightarrow \tau^- e^+) < 2.3 \times 10^{-9}
   \label{bound_B0}
\end{equation}
to be compared with the experimental limit \cite{pdg2010} of
$ Br(B^0 \rightarrow \tau^- e^+) < 2.8 \times 10^{-5}$.
Finally, for the decay $B^0_s \rightarrow \tau^- \mu^+$, using a $m_{B^0_s} = 5.366$ GeV \cite{pdg2010}
and $f_{B^0_s} = 0.24$ GeV \cite{Mohanta2010}, the bound
(\ref{bound_tramado}) gives a
\begin{equation}
  Br(B^0_s \rightarrow \tau^- \mu^+) < 2.7 \times 10^{-9}  \, .
  \label{bound_B0s}
\end{equation}
Unfortunately, for this decay there is no experimental information.
The $D^0$ and $B^0$ branching ratios are at least two orders of magnitude smaller than the
experimental upper bounds.
For $B^0_s$, the
experimental bounds coming from $g-2$ and muon decay predicts a branching ratio of the same order of magnitude as
for $B^0$. Note that, in the standard model, the decays discussed are not allowed at tree level but they are one-loop
allowed processes.
The same type of processes can give rise to the production of dark matter.
For example, the bounds (\ref{bound_D0}),  (\ref{bound_B0}), (\ref{bound_B0s}) also apply
to the decays where the leptons are replaced by their dark counter parts. These bounds suggests
that the branching rates for production of dark matter from $D$, $B^0$ and $B^0_s$ decays are, at most,
of the order of  $10^{-9}$. Note that the width are proportional to the lepton mass squared and vanish
for massless particles in the final state.

%==================================================================
%==================================================================
\section{WIMG Properties and WIMG Induced Processes}

%
%
%\textit{WIMG Width} -
%
%
The estimates for the WIMG mass suggest a $M \geq 9$ TeV. Then, from the point of view of the WIMG all
the particles in the multiplets (\ref{matter}) are massless. This simplifies considerably the computation of
the WIMG width and it follows
\begin{equation}
   \Gamma = \frac{g^2_M \, M}{24 \pi} \, N_F \, ,
   \label{gamma_1_8}
\end{equation}
where $N_F$ is the number of multiplets.
For the chiral theory (\ref{gamma_1_8}) should be multiplied by $1/2$. The bound (\ref{bound_tramado})
gives
$  \Gamma \leq 1.5 \times 10^{-5} \, M^3 \, N_F$,
where $\Gamma$ and $M$ are given in TeV. It follows that $\Gamma \approx 1$ TeV or smaller and, therefore,
the WIMG should have a very short lifetime $\tau = 1 / \Gamma \approx 7 \times 10^{-28}$ s.
This means that in the cosmic rays either the WIMG is produced via 
high energy processes or it is absent from the cosmic rays spectrum.

%==================================================================
%==================================================================
\subsection{A Proposal for WIMG Detection} 

\begin{figure}[t]
\centering
\epsfig{file=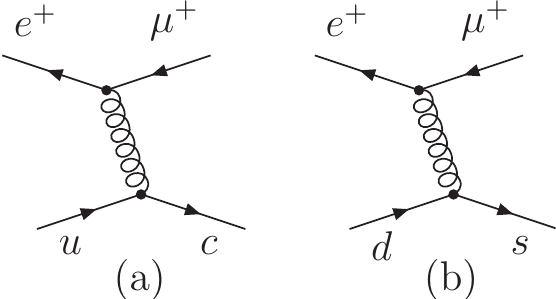,angle=0, width=6cm}
\vspace{.5cm}
\caption{Lepton-quark scattering with violation of the lepton family number and flavor exchange by WIMG
mediated processes.
Positron conversion to antimuon and flavor exchange $u\to c$ (a) and $d \to s$ (b). } \label{flavorx}
\end{figure}

The WIMG interaction can give rise to processes which are forbidden in the Standard Model. A window
to the detection of the new gauge boson is
the collision  $e^+ + p \rightarrow \mu^+ + X$, where $X = \Lambda$ or $\Lambda_c$, which
can occur via $t$-channel WIMG exchange but is forbidden
in the Standard Model. At the parton level, the tree-level
amplitude for lepton-quark scattering with violation of the lepton family number and flavor exchange is shown in
fig. \ref{flavorx} (a) and (b), for charm and strangeness production.
The total cross section at the parton level reads
\begin{eqnarray}
 \sigma (s) & = & \frac{1}{384 \, \pi} \, \frac{g^4_M}{M^4} \, s \, \left(1 - \frac{m^2}{s} \right)^2
 \, \left(2 + \frac{m^2}{s} \right)  \nonumber \\
 & \leq & 8.2 \times 10^{-13} \left( \frac{s}{1 \, \mbox{GeV}^2} \right)  \mbox{pbarn} ,
\end{eqnarray}
where $s$ is the c.m energy, $m$ the mass of the quark in the final state and the bound comes from
(\ref{bound_tramado}).
%So far, we have investigated the ordinary sector to constrain the parameters of
%the model through comparison with well established experimental results. The data clearly provide
%acceptable bounds for $g_M/M$ and set a scale for the WIMG lifetime.
%Experimental limits for the WIMP-nucleon interaction cross section, from
%precision recoil measurements on different nuclear targets, have been reported.
%Our WIMG model allows to compute this type of process involving dark and nuclear particles \cite{futuro}.

%==================================================================
%==================================================================
\subsection{Dark Proton--Ordinary Proton Cross Section} 

The dark sectors of the gauge model allow for the formation of dark nucleons which can interact with
ordinary matter via WIMG exchange. If the dark sectors mimic the ordinary matter sector and given the
current temperature of the Universe, the dark protons should be stable particles or, at least, their lifetime
should be of the same order of magnitude  as the lifetime of an ordinary proton. 

The cross section for the elastic scattering of a dark proton by an ordinary proton $\sigma (p_d p)$
is estimated from the one-WIMG exchange process as
\begin{equation}
  \sigma (p_d p) \approx  \frac{m^2_N}{64 \pi}  \left( \frac{g_M}{M} \right)^4 \, .
\end{equation}
For a nucleon mass of $m_N = 1$ GeV and using the bound (\ref{bound_tramado}), it follows that
$\sigma (p_d p) < 2.5 \times 10^{-48}$ cm$^2$.

This upper bound on $\sigma (p_d p)$ can be compared with the experimental bounds for the 
elastic cross section for the WIMP--nucleon scattering
$\sigma_{WIMP}$. The current experimental limits are summarized in, for example, 
Fig. 5 of \cite{XENON100}. For a WIMP mass of the order of a few GeV, the experimental limit gives
$\sigma_{WIMP} < 10^{-41}$ cm$^2$. The most stringent limit on the cross section occurs for a WIMP mass 
around 50 GeV,
where the corresponding lower limit on the WIMP--nucleon cross section is $7.0 \times 10^{-45}$ cm$^2$.

Our estimate of $\sigma (p_d p)$ gives a value which is several orders of magnitude below the experimental limit.

%==================================================================
%==================================================================
%\subsection{On the Abelian Portal}
%
%\begin{figure}[t]
%\centering
%\epsfig{file=fig4.eps,angle=0, width=6cm}
%\vspace{.5cm}
%\caption{Amplitude for photon--dark photon oscillation.} \label{abelian_portal}
%\end{figure}
%
%In the the gauge mirror model under discussion, the ordinary matter in the SM and the dark sectors interact
%via WIMG exchange. These couplings can induce oscillations between different sectors. 
%However, such oscillations are highly suppressed. For example, the model can accommodate an effective 
%interaction of the type explored in \cite{Demidov2012}  via the two loop diagram illustrated in Fig. 
%\ref{abelian_portal}. The diagram translates into an effective coupling between dark and ordinary photons of the
% type $F^{\mu\nu} F^\prime_{\mu\nu}$, where $F_{\mu\nu}$ is the Maxwell tensor for
%the electromagnetic field and $F^\prime_{\mu\nu}$ the corresponding tensor for the dark photon. 

%==================================================================
%==================================================================
\section{Conclusions} 

In the present work a mirror gauge model for the particle interactions between DM and ordinary matter
is developed. The model postulates a new SU(3) local symmetry, assumes the existence of various 
sectors, one being our ordinary matter, which are connected via the exchange of a new gauge boson. 

The model is compatible with Big Bang Nucleosynthesis and the recent measurements of the CMB. Further,
the dark sectors can be made collisionless if the temperature of the dark sectors is sufficiently lower than 
the observed temperature of the visible universe. This difference in the temperature seems to point
towards an asymmetric dark matter model, i.e. with the dark sectors not being exact copies of the
SM sector. 

The model is also compatible with particle physics phenomenology. The well established experimental 
results on the muon $\beta$-decay set a strong constraint on the model parameters. 
The data clearly provides bounds for $g_M/M$ and set a scale for the WIMG lifetime. 
Our estimate for the WIMG mass gives a value of the order of the electroweak scale
or larger, and a WIMG width of about 1 TeV. 
Such a large $\Gamma$ means that the WIMG should not be present in the cosmic rays spectrum unless
is produced via the collision of the fundamental particles.
Further, we have also checked that the contributions from WIMG exchange to FCNC in the ordinary sector
either vanish or are well below the current experimental limits. 

The new gauge boson can be associated with processes  that are forbidden in the SM. 
These type of reactions allows for the 
detection of the new $SU(3)$ symmetry with ordinary matter high energy experiments. 
As an example of SM forbidden
processes we suggest the reaction $e^+ + p \rightarrow \mu^+ + X$ where $X = \Lambda$ or $\Lambda_c$.
% at LHC and in cosmic rays \cite{futuro}.

Finally, the elastic cross for the scattering of an ordinary proton by a dark proton is estimated and compared
with the experimental upper bounds for the WIMP--proton cross section
\cite{DAMA2010,DAMA2010b,CDMS,CoGeNT2011a,CoGeNT2011b,CRESSTII2011,Wagner2011,XENON100,
XENON10,ZEPLIN2011}. The numbers provided combining our
estimate with the most demanding bounds coming from the muon beta decay give a
$\sigma (p_d p)$ which is several orders of magnitude lower than the upper bound on the
WIMP--proton cross section.
Our ordinary and dark matter model allows to compute this type of process involving dark and nuclear particles. 
In particular, we are currently involved in the calculation of the elastic and inelastic cross sections
of dark particles by nucleus which we plan to address in a future work \cite{futuro}.

%====================================================
%====================================================
\section*{Acknowledgements}

The authors acknowledge financial support from the Brazilian
agencies FAPESP (Funda\c c\~ao de Amparo \`a Pesquisa do Estado de
S\~ao Paulo) and CNPq (Conselho Nacional de Desenvolvimento
Cient\'ifico e Tecnol\'ogico) and the US Department of Energy
Grants DE-FG02-08ER41533, DE-SC0004971, DE- FC02-07ER41457 (UNEDF, SciDAC-2) and
the Research Corporation. OO acknowledges financial support from FCT under
contract PTDC/\-FIS/100968/2008.


\begin{thebibliography}{99}

\bibitem{WMAP1}
D. N. Spergel \textit{et al}., Astrophys. J. Suppl. \textbf{148} (2003) 175.

G. Hainshaw \textit{et al}., Astrophys. J. Suppl. Ser. \textbf{170} (2007) 288;
L. Pagel \textit{et al}., Astrophys. J. Suppl. Ser. \textbf{170} (2007) 355;
D. N. Spergel \textit{et al}., Astrophys. J. Suppl. \textbf{170} (2007) 377;
D. Larson \textit{et al}. (WMAP Collaboration), Astrophys. J. \textbf{192} (2011) 16.

\bibitem{Li2011}
M. Li, X.-D. Li, S. Wang, Y. Wang, Commun. Theor. Phys. \textbf{56}, 525 (2011) [arXiv:1103.5870].

\bibitem{Feng2010}
J. L. Feng, Ann. Rev. Astron. Astrophys. \textbf{48}, 495 (2010) [arXiv:1003.0904].

\bibitem{bertone05}
G. Bertone, D. Hopper, J. Silk,   Phys. Rep. \textbf{405} (2005) 279.

\bibitem{bertone10}
 G. Bertone, Nature \textbf{468} (2010) 389.
 
\bibitem{Pato2015} 
M.l Pato, F. Locco, G. Bertone, arXiv:1504.06324

\bibitem{Mavromatos2011}
N. E. Mavromatos, arXiv:1111.1563.

\bibitem{Arina2011}
C. Arina, J. Hamann, R. Trotta, Y. Y. Y. Wong, JCAP \textbf{3}, 8 (2012) [arXiv:1111.3238].

\bibitem{Fornengo2011}
N. Fornengo, P. Panci, M. Regis, Phys. Rev. \textbf{D84}, 115002 (2011) [arXiv:1108.4661].

\bibitem{DAMA2010}
R. Bernabei \textit{et al}. (DAMA/LIBRA Collaboration), Eur. Phys. J. \textbf{C56} (2008) 333.

\bibitem{DAMA2010b}
R. Bernabei \textit{et al}. (DAMA/LIBRA Collaboration), Eur. Phys. J. \textbf{C67} (2010) 39.

\bibitem{CDMS}
Z. Ahmed \textit{et al}. (CDMS Collaboration), Phys. Rev. Lett. \textbf{106} (2011) 131302.

\bibitem{CoGeNT2011a}
C. E. Aalseth \textit{et al}. (CoGeNT Collaboration), Phys. Rev. Lett. \textbf{106} (2011) 131301.

\bibitem{CoGeNT2011b}
C. E. Aalseth \textit{et al}. (CoGeNT Collaboration), Phys. Rev. Lett. \textbf{107} (2011) 141301.

\bibitem{CRESSTII2011}
G. Angloher \textit{et al}. (CRESST-II Collaboration), Eur. Phys. J. \textbf{C72}, 1971 (2012) [arXiv:1109.0702].

\bibitem{Wagner2011}
A. Drlica-Wagner \textit{et al}. (Fermi-LAT Collaboration), arXiv:1111.3358.

\bibitem{XENON100}
E. Aprile \textit{et al}. (XENON100 Collaboration), Phys. Rev. Lett. \textbf{105}  (2010) 131302.

\bibitem{XENON10}
J. Angle \textit{et al}. (XENON10 Collaboration), Phys. Rev. Lett. \textbf{107} (2011) 051301.

\bibitem{ZEPLIN2011}
D. Yu. Akimov (ZEPLIN-III Collaboration), Phys. Lett. \textbf{B 709}, 14 (2012) [arXiv:1110.4769].

\bibitem{Buen2015}
M. A. Buen-Abad, G Marques-Tavares, Martin Schmaltz, Phys. Rev. \textbf{D 92}, 023531 (2015).

\bibitem{Massey2015}
R. Massey \textit{et al}., MNRAS \textbf{449}, 3393 (2015).

\bibitem{Yang}
T.D. Lee and C. N Yang, Phys. Rev. \textbf{104} (1956) 254.

\bibitem{Kobzarev}
Y. Kobzarev, L. Okun, I. Pomeranchuk, Yad. Fiz. \textbf{3} (1966) 1154 .

\bibitem{Pavsic}
M. Pavsic, Int. J. Theor. Phys. \textbf{9} (1974) 229.

\bibitem{Foot}
R. Foot, H. Lew, R. Volkas, Phys. Lett. \textbf{B272} (1991) 67.

\bibitem{Akhmedov}
E. Akhmedov, Z. Berezhiani, G. Senjanovic, Phys. Rev. Lett. \textbf{69}, 3013 (1992).

\bibitem{futuro}
O. Oliveira, C. A. Bertulani, M. S. Hussein, W. de Paula, T. Frederico, in preparation.

\bibitem{Oliveira2011}
O. Oliveira, W. de Paula, T. Frederico, arXiv:1105.4899.

\bibitem{Berezhiani1996}
Z.G. Berezhiani, A.D. Dolgov, R.N. Mohapatra, Phys. Lett. \textbf{B375} (1996) 26.

\bibitem{Hoyle}
F. Hoyle and R.J. Tayler, Nature \textbf{203} (1964) 1108.

\bibitem{Kolb}
E. Kolb, D. Seckel, M. Turner, Nature \textbf{514} (1985) 415.

\bibitem{Berezhiani2001}
Z. Berezhiani, D. Comellic and F.L. Villante, Phys. Lett. \textbf{B503} (2001) 362.

\bibitem{Ciarcelluti}
P. Ciarcelluti, AIP Conf.Proc. \textbf{1038}, 202 (2008) [arXiv:0809.0668].

\bibitem{BBNLimit}
E. Lisi, S. Sarkar and F. L. Villante, Phys. Rev. \textbf{D59} (1999) 123520.

\bibitem{pdg2010}
K.A. Olive et al. (Particle Data Group), Chin. Phys. \textbf{C 38}, 090001 (2014)
%K. Nakamura \textit{et al.} (Particle Data Group), J. Phys. \textbf{G37}  (2010) 075021.


\bibitem{WMAP-cosm}
E. Komatsu \textit{et al}. (WMAP Collaboration), Ap.J. \textbf{192} (2011) 18.

\bibitem{ADM1}
H. Davoudiasl, D. E. Morrissey, K. Sigurdson, and S. Tulin, Phys. Rev. Lett. \textbf{105} (2010) 211304.

\bibitem{ADM2}
T. Cohen, D. J. Phalen, A. Pierce, and K. M. Zurek, Phys. Rev. \textbf{D 82}, 056001 (2010) [arXiv:1005.1655].

\bibitem{ADM3}
J. Shelton, K. M. Zurek, Phys. Rev. \textbf{D 82}, 123512 (2010)  [arXiv:1008.1997].

\bibitem{ADM4}
M. R. Buckley, L. Randall, 	JHEP \textbf{09} (2011) 009.

\bibitem{Mohapatra2002}
R. N. Mohapatra, S. Nussinov, V. L. Teplitz, Phys. Rev. \textbf{66} (2002) 063002.

\bibitem{Ackerman2009}
L. Ackerman, M. R. Buckley, S. M. Carroll, M. Kamionkowski, Phys. Rev. \textbf{79} (2009) 023519.

\bibitem{Mohanta2010}
R. Mohanta, Eur. Phys. J. \textbf{C71}, 1625 (2011) [arXiv:1011.4184].

\bibitem{Yoshida2000}
N. Yoshida, V. Springel, S. D. M. White, G. Tormen, Astrophys. J. \textbf{544} (2000) L87.

\bibitem{Blinnikov}
S. Blinnikov, M. Khlopov, Astron. Zh. \textbf{60} (1983) 632.

\bibitem{CiarcellutiStar}
Z. Berezhiani, P. Ciarcelluti, S. Cassisi, A. Pietrinferni, Astropart. Phys. \textbf{24} 6 (2006) 495.

\bibitem{Wandelt2000}
B. D. Wandelt, R. Dave, G. R. Farrar, P. C. McGuire, D. N. Spergel, P. J. Steinhardt, astro-ph/0006344.

\bibitem{Escude2000}
J. Miralda-Escude, Astrophys.J. \textbf{564}, 60 (2002)  [astro-ph/0002050].

\bibitem{Dave2001}
R. Dave, D. N. Spergel, P. J. Steinhardt, B. D. Wandelt, Astrophys. J. \textbf{547} (2001) 574.

\bibitem{Demidov2012}
S. V. Demidov, D. S. Gorbunov, A. A. Tokareva, Phys. Rev. \textbf{D 85} (2012) 015022.

\bibitem{Pavan2000}
M. M. Pavan, R. A. Arndt, I. I. Strakovsky, R. L. Workman, Phys. Scr. \textbf{T87} (2000) 65 [nucl-th/9910040].
%\bibitem{Antchev2011}
%G. Antchev \textit{et al}. (TOTEM Collaboration) arXiv:1110.1395.
\end{thebibliography}
\end{document}